\documentclass[useAMS,traditabstract]{aa}
\usepackage{natbib}
\usepackage{txfonts}
\usepackage{aas_macros}
\usepackage{graphicx}
\graphicspath{{/murphy/seb_stuff/XTE/IGRJ17544_Oct2011/PAPER_CORRECTIONS/}}
\usepackage{fixltx2e}

\begin{document}
\title{X-ray pulsations from the region of the Supergiant Fast X-ray Transient IGR J17544$-$2619}
\author{S. ~P. ~Drave\thanks{sd805@soton.ac.uk}\inst{\ref{inst1}}\and A. ~J. ~Bird\inst{\ref{inst1}}\and L. ~J. ~Townsend\inst{\ref{inst1}}\and A. ~B. ~Hill\inst{\ref{inst1}}\and V. ~A. ~McBride\inst{\ref{inst1},\ref{inst2},\ref{inst3}}\and V. ~Sguera\inst{\ref{inst4},\ref{inst5}}\and A. ~Bazzano\inst{\ref{inst5}}\and D. ~J. ~Clark\inst{\ref{inst6},\ref{inst7}}}

\institute{School of Physics and Astronomy, University of Southampton, University Road, Southampton, SO17 1BJ, UK\label{inst1}
\and
Astronomy, Gravity and Cosmology Centre, Department of Astronomy, University of Cape Town, Rondebosch, 7701, South Africa\label{inst2}
\and
South African Astronomical Observatory, PO Box 9, Observatory, 7935, South Africa\label{inst3}
\and
INAF-IASF, Istituto di Astrofisica Spaziale e Fisica Cosmica, Via Gobetti 101, Bologna, Italy\label{inst4}
\and
INAF-IASF, Istituto di Astrofisica Spaziale e Fisica Cosmica, Via del Fosso del Cavaliere 100, 00133 Roma, Italy\label{inst5}
\and
Centre d'Etude Spatiale des Rayonnements, CNRS/UPS, BP 4346, 31028 Toulouse, France\label{inst6}
\and
CREATEC, Unit 8, Derwent Mill Commercial Park, Cockermouth, Cumbria, CA13 0HT, UK\label{inst7}
}
        
\date{Recieved $-$ 25/08/2011 / Accepted $-$ 06/01/2012}

\abstract{Phase-targeted \emph{RXTE} observations have allowed us to detect a transient 71.49$\pm$0.02\,s signal that is most likely to be originating from the supergiant fast X-ray transient IGR J17544$-$2619. The phase-folded light curve  shows a possible double-peaked structure with a pulsed flux of $\sim$4.8$\times$10$^{-12}$\,erg cm$^{-2}$ s$^{-1}$ (3$-$10\,keV). Assuming the signal to indicate the spin period of the neutron star in the system, the provisional location of IGR J17544$-$2619 on the Corbet diagram places the system within the classical wind-fed supergiant XRB region. Such a result illustrates the growing trend of supergiant fast X-ray transients to span across both of the original classes of HMXB in P$_{orb}$ $-$ P$_{spin}$ space.}


\titlerunning{X-ray pulsations from the region of the SFXT IGR J17544$-$2619}
\authorrunning{Drave et al.}
\maketitle

\section{Introduction}

Supergiant Fast X-ray Transients (SFXTs) are a new class of high mass X-ray binary (HMXB) system that have been unveiled over the lifetime of the \emph{INTEGRAL} mission \citep{Winkler2003}. These sources are characterised by rapid X-ray outbursts, with durations of the order of tens of minutes to tens of hours \citep{2005A&A...444..221S}, and an association with OB supergiant companion stars \citep{2006ESASP.604..165N}. To date there are 10 confirmed SFXTs clustered along the Galactic plane (see individual papers for details: \citealt{2006A&A...455..653P}, \citealt{2006ApJ...638..982N}, \citealt{2007A&A...469L...5R}, \citealt{2008A&A...482..113M}, \citealt{2008A&A...486..911N}, \citealt{2008A&A...492..163R}, \citealt{2009A&A...494.1013Z}, \citealt{2009MNRAS.392...45R}). There are also several candidate SFXTs which exhibit the same X-ray flaring behaviour but for which an optical/IR counterpart is yet to be determined (e.g. \citealt{2006ApJ...646..452S}). With the determination of source distances from optical/IR spectroscopy \citep{2008A&A...484..801R}, peak outburst luminosities of 10$^{36}$ $-$ 10$^{37}$\,erg s$^{-1}$ have been deduced \citep{2004ATel..252....1G}. Combined with the observation of quiescence states at 10$^{32}$\,erg s$^{-1}$ (\citealt{2008ATel.1493....1B}, \citealt{2005A&A...441L...1I}) this illustrates a very high X-ray dynamic range of 10$^{4}$$-$10$^{5}$ within these systems. X-ray pulsations have been detected in four confirmed SFXTs implying there are accreting neutron stars in these systems (for individual sources see: \citealt{2005A&A...444..821L}, \citealt{2007A&A...467..249S}, \citealt{2007ATel..997....1S}, \citealt{2008ApJ...687.1230S}).

\object{IGR J17544$-$2619} was first discovered as a hard X-ray transient source on 2003 September 17 \citep{2003ATel..190....1S} with the IBIS/ISGRI (\citealt{2003A&A...411L.131U}/\citealt{2003A&A...411L.141L}) instrument aboard \emph{INTEGRAL}. Two short outbursts, 2 and 8 hours long respectively, were observed on the same day, indicating fast and recurrent transient behaviour. A subsequent detection on 2004 March 8 \citep{2004ATel..252....1G} further illustrated the recurrent nature of the X-ray outbursts in IGR J17544$-$2619. After the \emph{INTEGRAL} detection,  IGR J17544$-$2619 was associated with the soft X-ray source \object{1RXS J175428.3-262035} (\citealt{2003ATel..191....1W}, \citealt{2000IAUC.7432....3V}) and discovered in archival \emph{Beppo-SAX} \citep{2004ESASP.552..427I} and \emph{XMM-Newton} data \citep{2004A&A...420..589G}. A \emph{Chandra} observation \citep{2005A&A...441L...1I} precisely located the source with a positional accuracy of 0.6'' (RA = 17:54:25.284, DEC = -26:19:52.62, J2000.0), confirming the  association of IGR J17544$-$2619 with \object{2MASS J17542527-2619526}. \citet{2006A&A...455..653P} classified the companion as an O9Ib star with a mass of 25$-$28\,M$_{\odot}$ at a distance of 2$-$4\,kpc.  Subsequently \citet{2008A&A...484..801R} performed SED fitting to the mid-IR spectrum and refined the distance estimate to the system as $\sim$3.6\,kpc.

Using long baseline IBIS/ISGRI light curves, \citet{2009MNRAS.399L.113C} identified the orbital period of IGR J17544$-$2619 as 4.926 $\pm$ 0.001\,d, one of the shortest orbital periods observed in an SFXT.

The spectral properties of IGR J17544$-$2619 have been studied at all levels of emission.  The outburst spectra are often well fit with powerlaw models that show variations in column density, 1.1$-$3.3$\times$10$^{22}$\,cm$^{-2}$, and photon index, 0.75$-$1.3, between outbursts (\citealt{2009ApJ...707..243R}, \citealt{2009ApJ...690..120S}, \citealt{2008ATel.1697....1R}). The quiescence spectra are far softer than those of outbursts with photon indices between 2.1 and 5.9 (\citealt{2008ApJ...687.1230S}, \citealt{2005A&A...441L...1I}). The softness of these spectra has led to the conclusion that the compact object in IGR J17544$-$2619 is most likely a neutron star (\citealt{2005A&A...441L...1I}, \citealt{2006A&A...455..653P}). In this paper we outline a new study performed on the IGR J17544$-$2619 system. Our data set is described in Sect. \ref{dsetanda}, followed by temporal and spectral results in Sect. \ref{results}. A discussion of these results is given in Sect. \ref{discuss} followed by conclusions in Sect. \ref{conc}.

\label{intro}

\section{Data Set and Analysis}

Using the orbital ephemeris of \citet{2009MNRAS.399L.113C} three observations of the region around IGR J17544$-$2619 were performed through one half of the compact object orbit using the Proportional Counter Array (PCA) instrument aboard \emph{RXTE} (\citealt{2006ApJS..163..401J}, \citealt{1994AAS...185.6701S} respectively). The observations were performed on 2010 May 15, 16 and 17 at 04:47, 16:54 and 03:53 UTC with each observation having an exposure of $\sim$ 10\,ks. In all subsequent analysis we now use an updated orbital period determination and periastron ephemeris of 4.9278$\pm$0.0002\,d and MJD 53732.632 respectively, placing periastron at an orbital phase of 0.0. These improved values were determined through utilising the analysis methods outlined in \citet{2009MNRAS.399L.113C} with an \emph{INTEGRAL}/IBIS data set of 13.3\,Ms, representing an $\sim$66\% increase in exposure compared with the original study. The orbital phase of the three observations are then 0.412, 0.720 and 0.811 respectively.

The data were analysed using the standard tools within HEASOFT v6.9. The analysis was performed on both the science array (standard mode one) and science event (good xenon) data. PCU 2 was the only detector active throughout all three observations, with PCU 4 active for a small percentage of each, hence extraction was performed on PCU 2 data only. The GTIs of the observations were generated with the ftool MAKETIME using the observation filter files and the standard event selection criteria described in the \emph{RXTE} data analysis manual. Light curves of the standard mode one data, which covers the full energy response of the PCA at 0.125\,s time resolution, were extracted using the ftool SAEXTRCT. The light curves were background subtracted using the most recent PCA faint background model and barycentered using the FXBARY ftool. Similarly the event mode data were treated in the same manner, whereby the light curves were extracted using SEEXRCT at 0.125\,s resolution. Three light curves were extracted covering  the full PCA energy response, 3$-$10\,keV and 10$-$120\,keV with all layers of PCU 2 used in each case. Again the light curves were barycentered using FXBARY. The science event data was also used to create energy spectra of the observations. The spectra, background and response files were generated for each science event file and for every combination of active detectors using the standard methods. As the majority of the exposure in each observation was achieved whilst PCU 2 was the only active detector, these spectra, and backgrounds, were summed using SUMPHA and the responses combined using ADDRMF.     

\begin{figure}
	\includegraphics[scale=0.5]{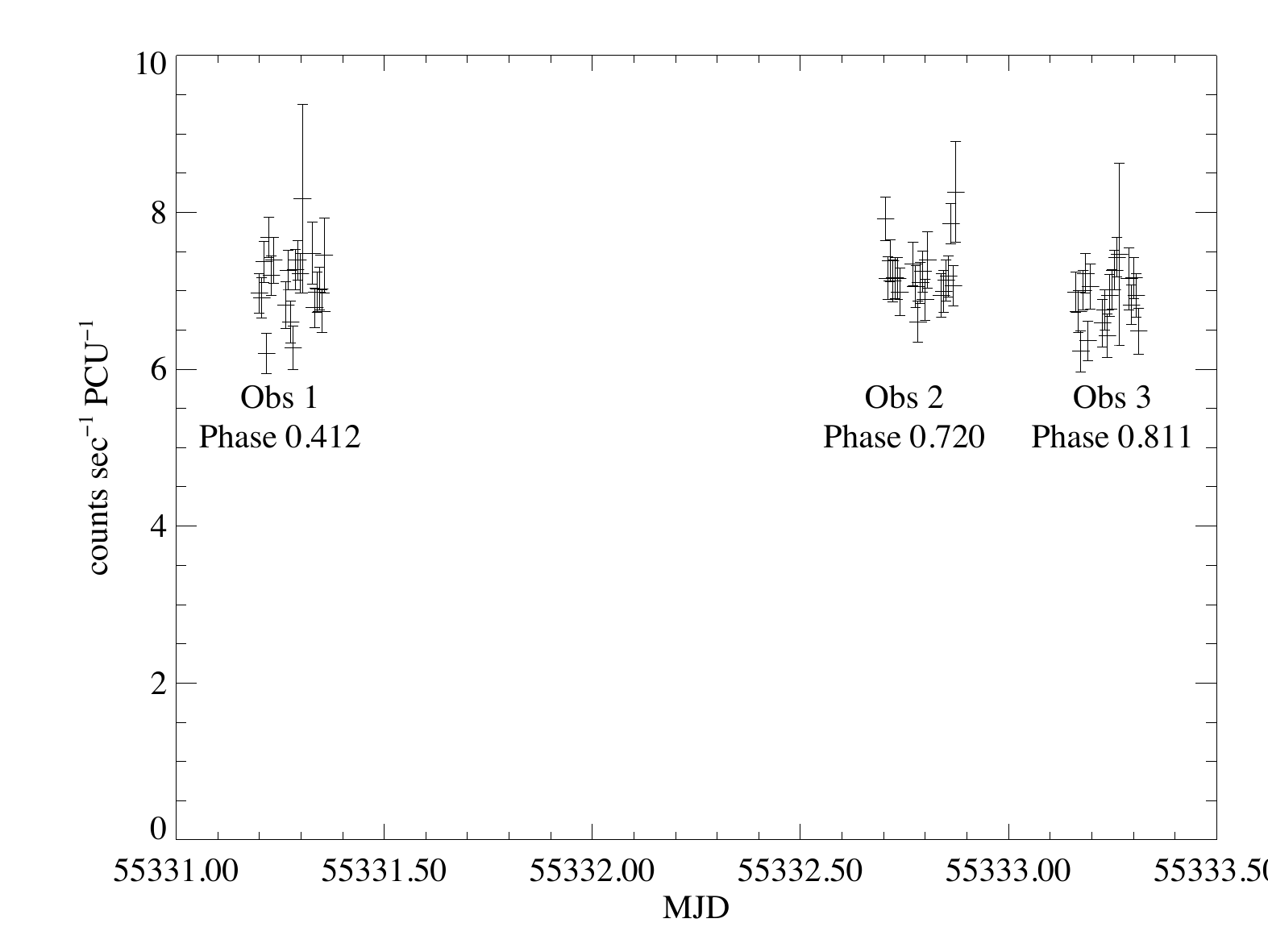}
	\caption{Background subtracted, binned standard mode one light curves of the \emph{RXTE}/PCA observations (bin time: 500\,s). The orbital phase of each observation is shown using the zero phase ephemeris MJD\,53732.632 and orbital period 4.9278\,d where periastron occurs at orbital phase 0.0.}
	\label{500sec_lightcurve}
\end{figure}

\label{dsetanda}

\section{Results}

Figure \ref{500sec_lightcurve} shows the PCA standard mode one light curves of the observations at 500\,s binning. In this mode PCA data has no energy resolution and hence these represent the emission detected across the full 2$-$120\,keV PCA energy range. We see that steady, low level emission at an intensity of $\sim$7 counts s$^{-1}$ PCU$^{-1}$ is being detected across all three observations. This corresponds to a flux of $\sim$ 5.0$-$5.5$\times$10$^{-11}$\,erg cm$^{-2 }$ s$^{-1}$ in the 3$-$10\,keV energy range using the spectral models outlined in Table \ref{spectralfits}. The orbital phase of the centre point of each observation is also indicated and it is seen that the first observation (Obs 1) occurs just before apastron whilst the second and third observations (Obs 2 and 3) were performed as the compact object is approaching periastron.

The shape and intensity of the emission in the three observations was first characterised by assessing the statistical properties of the finely binned (0.125\,s) standard mode one light curves. The mean fluxes are seen to vary by a maximum of $\sim$5\% between the observations. The level of variation within each observation was also investigated by means of the goodness of fit to a constant flux at the average count rate. All three observations showed a poor fit, indicating variations in excess of statistical noise within the light curves. These statistics are outlined in Table \ref{statsTable}. The properties of the observed emission suggest that there are no sources undergoing bright, fast outbursts within these observations.

 \begin{table}
 \caption{Statistical properties of the 0.125\,s resolution 2 $-$120\,keV standard mode one light curves.} 
 \begin{center}
 \begin{tabular}{c c c c}
  \hline
  \multicolumn{1}{c}{Obs} & \multicolumn{1}{c}{Average Flux} & \multicolumn{1}{c}{$\hat{\chi}^{2}$} & \multicolumn{1}{c}{Exposure} \\
   & counts s$^{-1}$ PCU$^{-1}$ &  & s \\ \hline
  1 & 7.02 $\pm$ 0.06 & 12.1 & 9102  \\
  2 & 7.19 $\pm$ 0.06 & 10.9 & 9722  \\
  3 & 6.87 $\pm$ 0.06 & 12.2 & 8872  \\ \hline
 \end{tabular}
 \end{center}
 \label{statsTable}
\end{table}

\label{results}

\subsection{Periodicity Analysis}

\begin{figure}
	\includegraphics[scale=0.5]{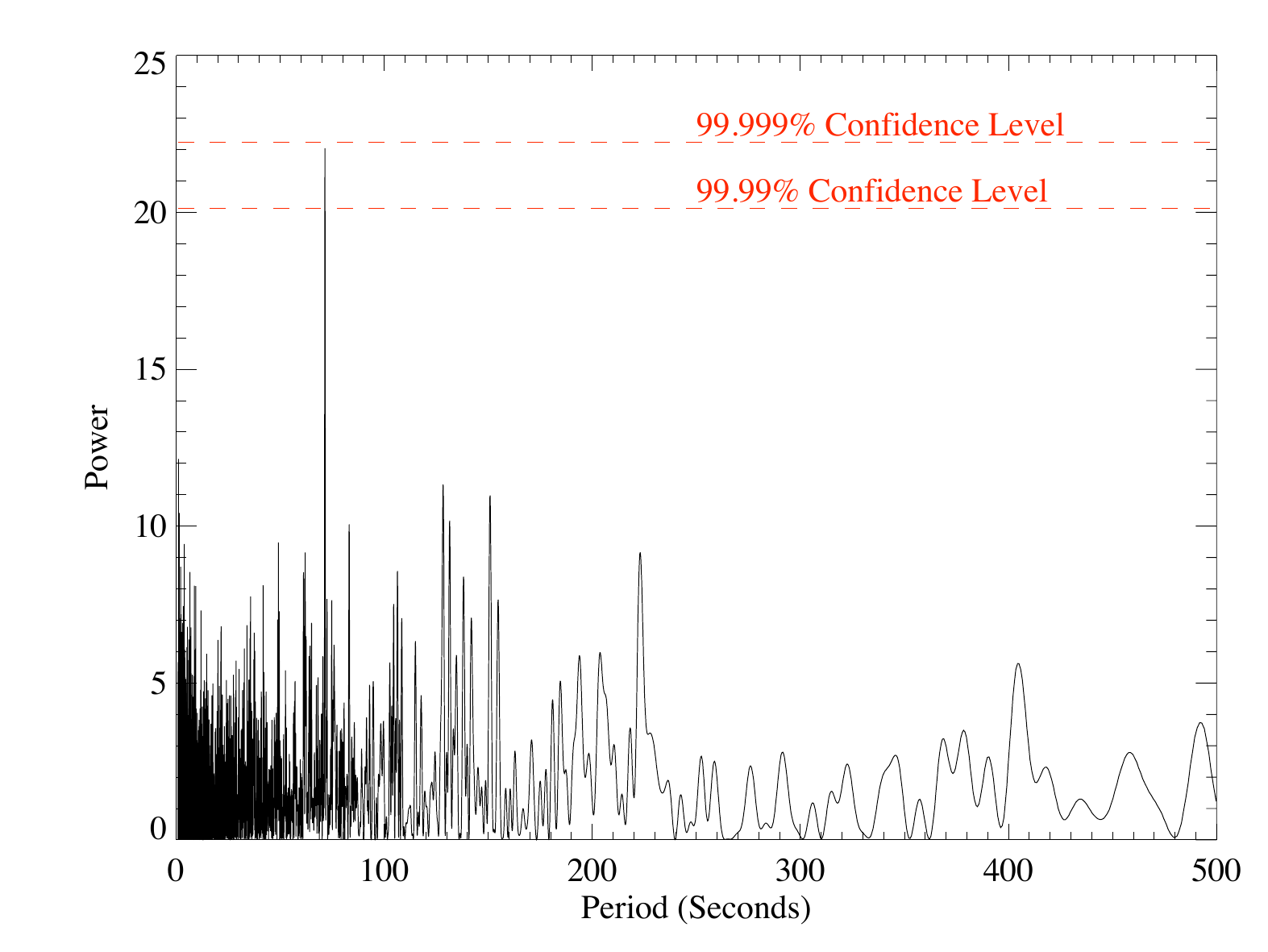}
	\caption{Lomb-Scargle periodogram of the background subtracted Obs 1 0.125\,s resolution standard mode one lightcurve showing a peak at a period of 71.49\,s. This period is calculated to have a significance of 4.37$\sigma$.}
	\label{LS_conf}
\end{figure}

\begin{figure}
	\includegraphics[scale=0.5]{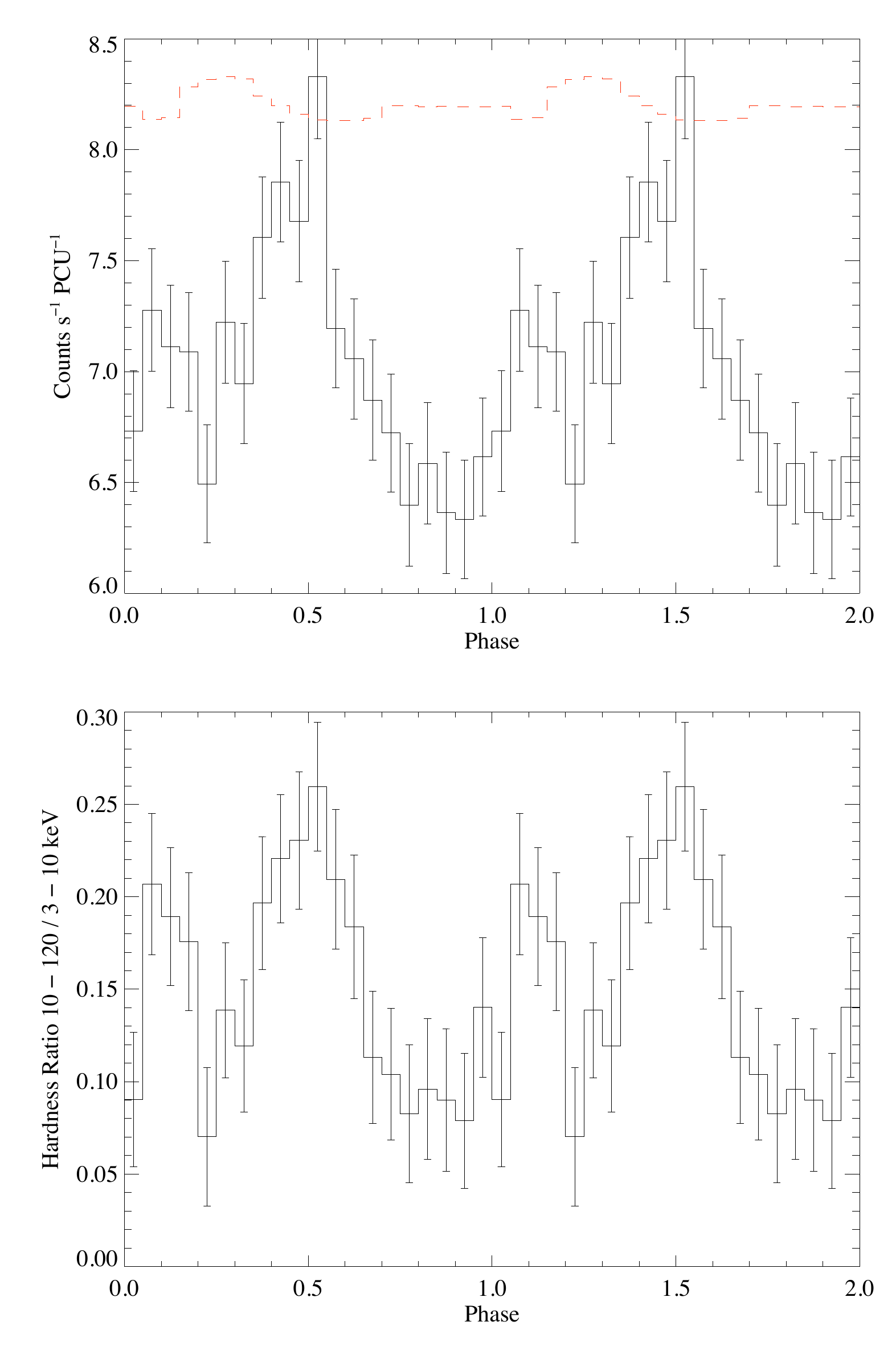}
	\caption{{\bf Top:} Obs 1 light curve phase-folded on the 71.50\,s period. The profile shows a pulse fraction of $\sim$13\%. The dashed line shows the relative exposure of each phase bin (a relative exposure of 1 equates to a count rate of 8.3 and 0 to 0 within this scaling). {\bf Bottom:} Hardness ratio between the 3$-$10 and 10$-$120\,keV energy bands showing a hardening of the observed emission during the peaks in the above phase folded light curve. Both curves posses the same phase binning and arbitrary zero pulse phase ephemeris.}
	\label{pfold}
\end{figure}

To search for pulsations in the data, the background subtracted, 0.125\,s resolution standard data mode one light curves of each observation were subjected to a Lomb-Scargle analysis (\citealt{1976Ap&SS..39..447L}, \citealt{1982ApJ...263..835S}). The analysis of Obs 2 and 3 showed no significant signals within their periodograms. However, a significant peak at 71.49\,s was observed with a power of 22.039 in the periodogram of Obs 1, as shown in Fig. \ref{LS_conf}. This peak does not correspond to a beat frequency between any characteristic time scales within the data and is present when the analysis is performed on light curves with a wide range of different time binnings. The 99.99\% and 99.999\% confidence levels were calculated at Lomb-Scargle powers of 20.133 and 22.227 respectively using a randomisation test as outlined in  \citealt{2005A&A...439..255H} whilst also taking into account the extra trials resulting from the analysis of the light curves from the three observations. By linearly interpolating between these confidence levels we calculate the significance of the 71.49\,s period as 4.37$\sigma$. A similar randomisation based test was used to estimate the error on this period as $\pm$0.02\,s (see \citealt{2010MNRAS.409.1220D} for details). Figure \ref{pfold} shows the phase-folded light curve of Obs 1 using the 71.49\,s period. The shape is dominated by a large peak on top of an underlying flux at $\sim$6.4\,counts s$^{-1}$ PCU$^{-1}$, representing a pulse fraction of $\sim$13\%, where the pulse fraction is defined as (C$_{max}$ $-$ C$_{min}$)/(C$_{max}$ $+$ C$_{min}$). A possible second peak is also observed at lower significance and is offset from the larger peak by a phase of $\sim$0.3$-$0.4. A discussion of this shape and the non-detections in Obs 2 and 3 is given in Sect. \ref{discuss}

For consistency the event mode data were also investigated for periodicities using the extracted 0.125\,s resolution background subtracted light curves. Using the full energy response light curve (2$-$120\,keV) the only significant feature in the periodogram is detected at 71.52\,s with a power of 20.37, in agreement with the result from the standard mode one data. The event mode data light curves in the 3$-$10 and 10$-$120\,keV energy bands were also searched for pulsations (Note: When using restricted energy ranges 3\,keV is used as the low energy cut-off point due to the degrading of the PCA energy calibration below this value). In both cases a peak at $\sim$71.5\,s is observed in the periodogram but it is not at a significant level. The sum of these periodograms does however, produce a single significant peak at the correct periodicity. We take this to illustrate that due to the faintness of the emission a significant detection can only be obtained when the full energy range is included in the data set and hence the periodicity is present (to some extent) over a broad range of energies.
 
Variations in the hardness ratio as a function of pulse phase have been investigated. Figure \ref{pfold} shows the 2$-$120\,keV phase-folded light curve (top) above the phase-folded 10$-$120 to 3$-$10\,keV hardness ratio (bottom). It can be seen that the emission hardens during the two pulse phase regions that are coincident with increased flux in the phase-folded light curve. This suggests a physical origin for both the high and lower significance peaks seen in the phase-folded light curve shown in Fig. \ref{pfold} (top).

\label{perA}

\subsection{Spectral Analysis}

The energy spectra derived from the total data in each observation were investigated to characterise the spectral shape of the emission and aid in the identification of its source. The spectra were fit with models in the 3$-$20\,keV energy range using XSPEC v12.4. All reported errors from the spectral fits are quoted at the 90\% confidence level. The spectra were initially fit with simple absorbed models (e.g. phabs(bremss)), however the large reduced Chi-Squared values (6.8$-$12.8 for 37 degrees of freedom) showed a need for an additional Gaussian component interpreted as an iron emission line. It was found that absorbed power-law models again with this additional Gaussian component, gave the best fits across all three observations. Power-laws with high energy cut-offs and thermal bremsstrahlung models were also found to produce statistically similar fits for some individual observations, but neither could produce good fits across all three observations.

The results of the power-law fits are outlined in Table \ref{spectralfits}. It is seen that the spectra are quite highly absorbed and do not show a significant variation in column density across the observations. They possess photon indices that vary from 2.37$^{+0.09}_{-0.09}$ to 2.69$^{+0.13}_{-0.13}$ between Obs 1 and 3. The inferred 3$-$10\,keV fluxes also show some variation across the three observations, declining significantly from 5.43$^{+0.08}_{-0.10}\times$10$^{-11}$ in Obs 1 to 5.13$^{+0.10}_{-0.12}\times$10$^{-11}$ erg cm$^{-2}$ s$^{-1}$ in Obs 3. The flux in Obs 2 lies between these two values, suggesting a decline throughout the observations. The spectra of the individual observations were also summed and the total spectrum fitted. The best fit parameters are shown Table \ref{spectralfits} and a similar shape is again observed.

The large equivalent width of the iron line component in each spectrum, varying from 0.78 to 0.88\,keV, is of interest. These values are consistent with those quoted as resulting from Galactic Ridge emission \citep{1986PASJ...38..121K}, suggesting that Galactic Ridge emission could account for part or all of the emission detected in these observations. To investigate this we fit the spectra with the model of the central ridge region presented in Table 3 of \citet{1998ApJ...505..134V}. The Raymond-Smith plasma temperature and power-law photon index were fixed to kT$=$2.9\,keV and $\Gamma =$1.8 respectively, whilst the absorption and normalisations were left as free parameters. As Table \ref{valinafits} shows, Obs 2 and 3 are well fit by this model. However Obs 1 is not well fit by this model, indicating that Galactic Ridge emission does not fully describe the spectral shape of this observation. A decreasing unabsorbed flux trend is also observed between Obs 1 and 3. Combining the poor fit given by the Galactic Ridge model in Obs 1 and the decreasing flux trend seen across the three observations with the fact that a pulsation is only seen during Obs 1, is taken to show that in this observation there is an additional source of X-ray emission that is generating the pulsed signal that is observed in addition to the Galactic Ridge emission.

To further investigate the possible nature of the additional X-ray source we used pulse phase resolved spectroscopy to extract spectra during the pulse on, phase 0.4$-$0.7 in the top panel of Fig. \ref{pfold}, and pulse off, 0.7$-$1.0, phase regions. The spectrum collected from the pulse off region was then subtracted from the pulse on region, this calculation was performed in `count rate' space to compensate for the different exposure times accumulated for each phase region.  However, whilst there were residual counts in the subtracted spectrum the signal-to-noise was not sufficient to allow the fitting of spectral models to characterise the pulsed emission and provide a direct estimate of the residual flux. Figure 3 (bottom) does show a hardening of the emission during the pulse on phase region however, suggesting the presence of an additional active, pulsing X-ray source within the FOV as the Galactic Ridge emission does not vary on these time scales.

We can make a refined estimate of the pulsed flux in Obs 1 by using the phase-folded light curve shown in the top panel of Fig. \ref{pfold}. Under the assumption that the minimum count rate in the phase-folded light curve corresponds to zero pulsed emission, which is supported by the consistency of the count rate between phases 0.70 and 1.0 in Fig. \ref{pfold}, we calculate the percentage excess above this minimum count rate in each phase bin. Taking the average of the excesses observed in each phase bin then produces an estimate of the pulsed flux fraction as 8.9\%. Using the Obs 1 flux value obtained from the spectral fits outlined in Table \ref{spectralfits} this corresponds to a flux of 4.8$\times$10$^{-12}$\,erg cm$^{-2}$ s$^{-1}$ (3$-$10\,keV) originating from the pulsed signal with the remainder resulting from the constant Galactic Ridge emission. As the Galactic Ridge emission cannot generate a pulsed signal or spectral variations on a time scale of tens of seconds we therefore attribute this pulsed flux component to another active X-ray source within the FOV during Obs 1.

\begin{table*}
 \begin{minipage}{\textwidth}
 \caption{Spectral fits to the total and three individual observations using the model: Phabs(Powerlaw + Gaussian). Error are quoted at the 90\% confidence level} 
 \begin{center}
 \begin{tabular}{c c c c c c c c}
  \hline
  \multicolumn{1}{c}{Obs} & \multicolumn{1}{c}{$\hat{\chi}^{2}$/d.o.f.} & \multicolumn{1}{c}{$\Gamma$} & \multicolumn{1}{c}{n$_{H}$} & \multicolumn{1}{c}{Line Energy} & \multicolumn{1}{c}{Line Sigma} & \multicolumn{1}{c}{Line Equivalent} & \multicolumn{1}{c}{Flux (3$-$10\,keV)} \\
   & & & 10$^{22}$ cm$^{-2}$ & keV & keV & Width keV & erg cm$^{-2}$ s$^{-1}$ \\ \hline\noalign{\smallskip}
  1 & 0.95/34 & 2.37$^{+0.09}_{-0.09}$ & 4.4$^{+1.3}_{-1.2}$ & 6.54$^{+0.05}_{-0.05}$ & 0.03$^{+0.19}_{-0.03}$ & 0.82 & 5.43$^{+0.08}_{-0.10}\times$10$^{-11}$  \\
   & & & & & & \\
  2 & 0.39/34 & 2.40$^{+0.13}_{-0.13}$ & 5.1$^{+1.7}_{-1.7}$ & 6.56$^{+0.07}_{-0.07}$ & 0.15$^{+0.18}_{-0.15}$ & 0.78 & 5.28$^{+0.10}_{-0.11}\times$10$^{-11}$  \\
   & & & & & & \\
  3 & 0.78/34 & 2.69$^{+0.13}_{-0.13}$ & 5.9$^{+1.6}_{-1.6}$ & 6.67$^{+0.07}_{-0.06}$ & 0.14$^{+0.17}_{-0.14}$ & 0.88 & 5.13$^{+0.10}_{-0.12}\times$10$^{-11}$  \\ \noalign{\smallskip}\hline\noalign{\smallskip}
  Total & 1.01/34 & 2.48$^{+0.06}_{-0.06}$ & 5.2$^{+0.8}_{-0.8}$ & 6.58$^{+0.03}_{-0.03}$ & 0.11$^{+0.11}_{-0.10}$ & 0.80 & 5.32$^{+0.06}_{-0.06}\times$10$^{-11}$ \\ \noalign{\smallskip}\hline
 \end{tabular}
 \label{spectralfits}
 \end{center}
 \end{minipage}
\end{table*}

 \begin{table}
 \caption{Spectral fits to the three individual observations using the model Phabs(Raymond + Powerlaw) of \citet{1998ApJ...505..134V} } 
 \begin{center}
 \begin{tabular}{c c c c c}
  \hline
  \multicolumn{1}{c}{Obs} &  \multicolumn{1}{c}{ } & \multicolumn{1}{c}{$\hat{\chi}^{2}$} &  \multicolumn{1}{c}{ } & \multicolumn{1}{c}{Flux (3$-$10\,keV)} \\
   & & &  & erg cm$^{-2}$ s$^{-1}$ \\ \hline\noalign{\smallskip}
  1 & & 1.735 & & 5.41$^{+0.09}_{-0.08}\times$10$^{-11}$  \\
   & & & & \\
  2 & & 0.700 & & 5.26$^{+0.10}_{-0.08}\times$10$^{-11}$  \\
   & & & & \\
  3 & & 0.869 & & 5.13$^{+0.07}_{-0.09}\times$10$^{-11}$  \\ \noalign{\smallskip}\hline
 \end{tabular}
 \end{center}
 \label{valinafits}
\end{table}

\label{specA}

\section{Discussion}

Following the spectral analysis presented in Sect. \ref{specA} we have concluded that the emission observed in Obs 2 and 3 is most likely originating from the diffuse Galactic Ridge emission. However in Obs 1 there appears to be evidence for an additional flux component that is generating a periodic signal. This emission is attributed to a further active X-ray source within the FOV during this observation. As no significant periodic signals were observed during Obs 2 and 3, and no structure was seen when these light curves were folded on the known 71.49\,s period, we conclude that the additional X-ray source was not active during these observations. This interpretation is supported by the significant decrease in 3 $-$ 10\,keV flux observed between Obs 1 and 3, see Sect. \ref{specA}. We note however that this interpretation represents the simplest situation whereby we only use two flux components that we can definitively identify, namely the pulsed flux in Obs 1 and a contribution from the Galactic Ridge emission in all three observations. In fact it may be that there are additional faint, non-pulsating sources within the PCA FOV during all three observations that contribute towards the detected flux in each. Changes in the flux emitted by or the number of any such sources could cause the variations in the average 2 $-$ 120\,keV flux detected in each observations as outlined in Table \ref{statsTable}. Due the non-imaging nature of the PCA however it is not possible to perform identification of any non-pulsating source other than the Galactic Ridge emission, identified by the large equivalent width of the iron line component in the spectral fits, and hence we use the most simplistic interpretation for the remainder of this paper. A second consequence of the non-imaging nature of the PCA is that we are also required to give further consideration as to the source of the excess, pulsed emission seen in Obs 1.

\begin{figure}
	\includegraphics[scale=0.45]{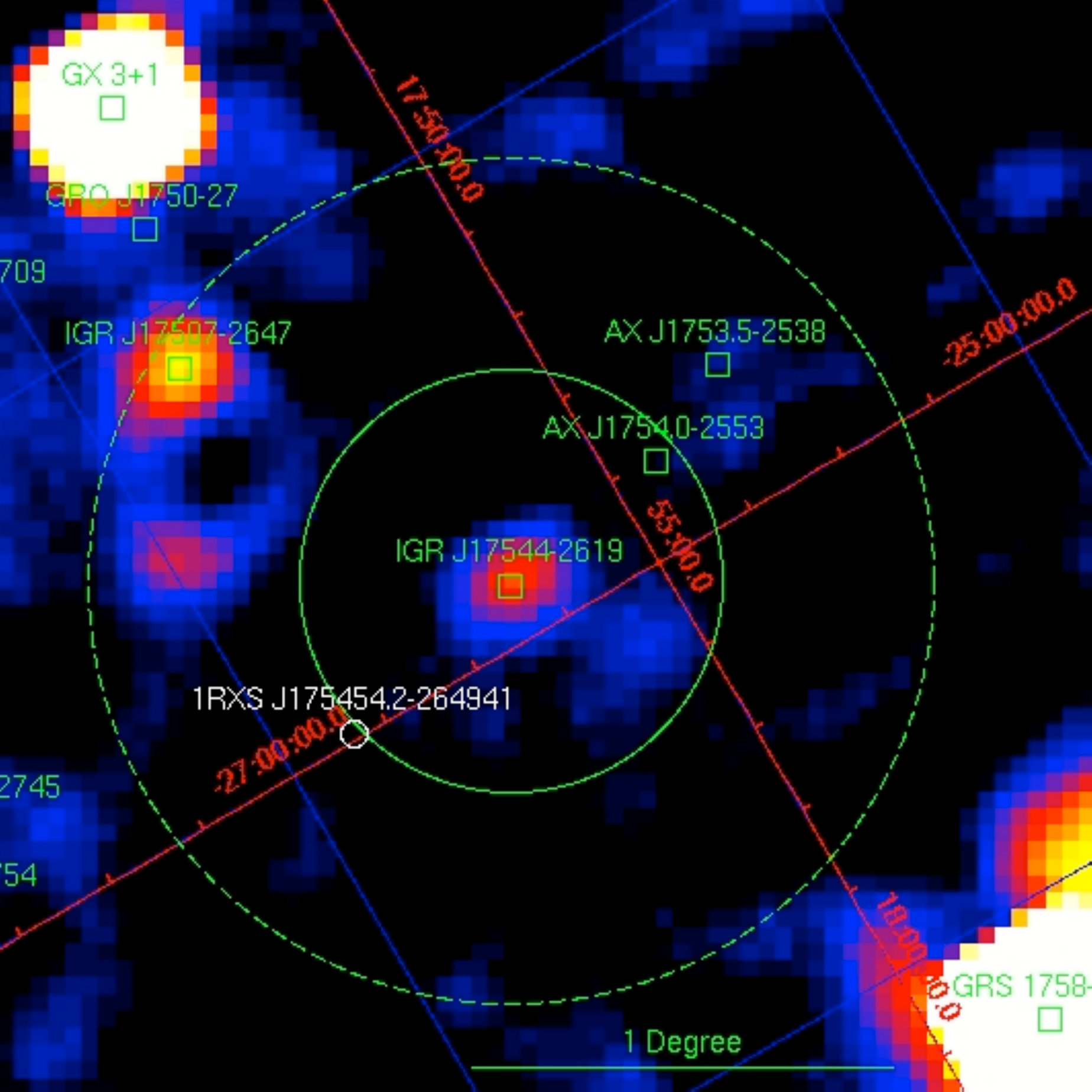}
	\caption{The IBIS survey 18$-$60\,keV significance map of the IGR J17544$-$2619 region, exposure $\sim$8\,Ms, with the PCA FOV half and zero response contours overlaid \citep{2010ApJS..186....1B}. The sources in this region that are contained in the \emph{INTEGRAL} general reference catalog are shown as square points, whilst further X-ray sources are shown as circles.}
	\label{cat4_map}
\end{figure}

\begin{figure}
	\includegraphics[scale=0.55]{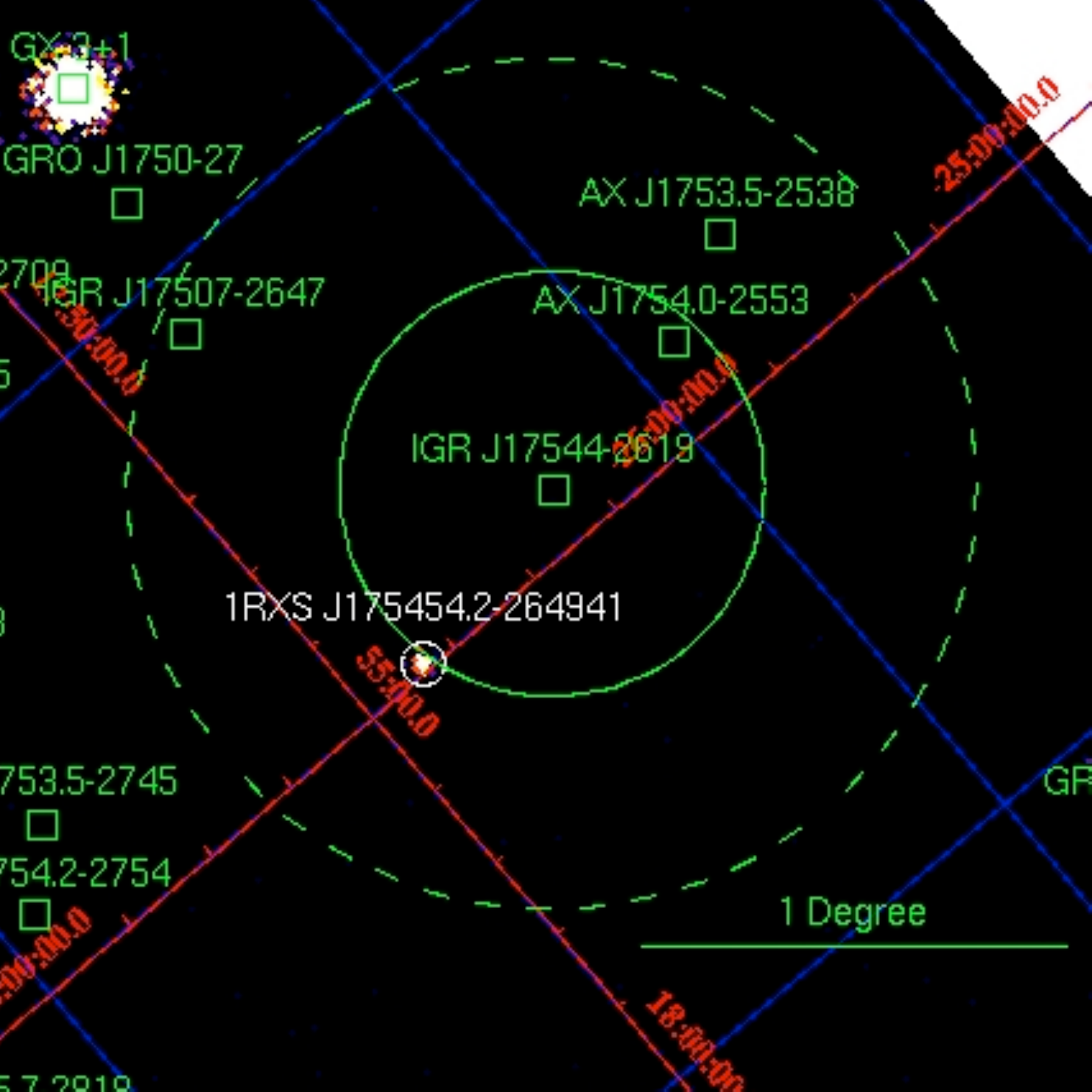}
	\caption{The ROSAT all sky survey photon map of the IGR J17544$-$2619 region in the 0.1$-$2.4\,keV energy range \citep{1999A&A...349..389V}. The annotations are the same as those in Fig. \ref{cat4_map}.}
	\label{ROSAT_map}
\end{figure}

Figure \ref{cat4_map} shows the fourth \emph{INTEGRAL}/IBIS survey significance map of the region in the 18$-$60\,keV energy band \citep{2010ApJS..186....1B}. Overlaid are the PCA half and zero collimator response contours, at 0.5$^{o}$ and 1$^{o}$ respectively, for the pointing used, along with the sources in the \emph{INTEGRAL} general reference catalog (squares) (v.31, \citealt{2003A&A...411L..59E}) and further X-ray detections (circle). Similarly the \emph{ROSAT} all sky survey photon map of the region in the 0.1$-$2.4\,keV energy range is shown in Fig. \ref{ROSAT_map} \citep{1999A&A...349..389V}. It can be seen that the only two sources significantly detected within the zero response contour by \emph{INTEGRAL} are IGR J17544$-$2619 and \object{IGR J17507$-$2647}. The latter of these sources, which is at the edge of the FOV, was characterised using \emph{Chandra} observations by \citet{2009ApJ...701..811T} as a weak, persistent source with a flux of 4.5$\times$10$^{-12}$\,erg cm$^{-2}$ s$^{-1}$ (0.2$-$10\,keV) that is most likely a distant HMXB at $\sim$8.5\,kpc. The \emph{Chandra} spectrum of the source reported showed a high level of absorption with n$_{H}$ $=$ 1.34$\times$10$^{23}$\,cm$^{-2}$ and the source was not detected in the \emph{ROSAT} map. As IGR J17507$-$2647 is located at the edge of the PCA FOV it is in a region of low collimator response ($<$ 5\%) and would therefore require a flux of at least a factor of 20 greater than that reported by \citealt{2009ApJ...701..811T} to generate the observed pulsed flux. As this is a persistent source with no reported outbursts we conclude that the emission observed by \emph{RXTE} is unlikely to be contaminated by IGR J17507$-$2647.

The soft X-ray source \object{1RXS J175454.2-264941} \citep{1999A&A...349..389V} is reported in the \emph{ROSAT} bright source catalog and is located near the half response contour; it is the only bright soft X-ray source detected within the PCA FOV by \emph{ROSAT}. The hardness ratio reported suggests the source is moderately hard (0.82 for the 0.5$-$2/0.1$-$0.4\,keV energy bands) and could be detected by the PCA. There are also four sources from the \emph{ROSAT} faint source catalog \citep{2000IAUC.7432....3V} within the FOV, however we would not expect a detection of these sources using the PCA. Finally the ASCA sources shown in Fig. \ref{cat4_map} are reported as X-ray point sources by \citet{2001ApJS..134...77S}. \object{AX J1753.5$-$2538} was detected at 3$\sigma$ in the 0.7$-$2\,keV band but was not found in the 2$-$10\,keV band indicating that it is a soft source, hence we would not expect a detection with PCA. However the non-detection in Fig. \ref{ROSAT_map} does suggest a transient nature for this source. \object{AX J1754.0$-$2553} is not detected in the 0.7$-$2\,keV band but does have a 3.8$\sigma$ detection in the 2$-$10\,keV energy range. This suggests the source is harder and, as for 1RXS J175454.2$-$264941, could be detected in the PCA data. 

Of all the sources discussed above, IGR J17544$-$2619 is the most active showing a large number of outbursts that have been detected by a large variety of missions \citep{2009MNRAS.399L.113C}. Taking into account the nature of the known sources within the PCA FOV makes IGR J17544$-$2619 the most likely known source of the emission detected by the PCA instrument. For the remainder of this paper we assume this to be the case, although we cannot rule out the possibility that the emission is coming from one of the other known sources, 1RXS J175454.2$-$264941 and AX J1754.0$-$2553 in particular, or a new unknown source within the FOV. 

If we consider the 71.50\,s signal as a pulsation from the IGR J17544$-$2619 system, then this confirms that the compact object in the system is a neutron star, as has been suggested from quiescence spectra (\citealt{2005A&A...441L...1I}, \citealt{2006A&A...455..653P}). Assuming a source distance of $\sim$3.6\,kpc this makes the estimated source flux equivalent to an unabsorbed luminosity of $\sim$1$\times$10$^{34}$\,erg s$^{-1}$ (3$-$10\,keV), indicating that IGR J17544$-$2619 was observed in a low X-ray state as opposed to during one of its large outbursts (such as the 10$^{36}$ erg s$^{-1}$ event reported in \citealt{2004ATel..252....1G}). X-ray pulsations have also been detected in other SFXT systems observed during similar low luminosity states, $\sim$1$\times$10$^{34}$\,erg s$^{-1}$ (0.5$-$10\,keV) in \object{IGR J18483$-$0311} \citep{2009MNRAS.399..744G} and 2.3$\times$10$^{34}$\,erg s$^{-1}$ (2$-$10\,keV) in \object{AX J1841.0-0536} \citep{2001PASJ...53.1179B} for example. Additionally the hardness ratio has also been seen to vary as a function of pulse phase in the SFXT IGR J11215$-$5952 \citep{2007A&A...476.1307S}, showing a hardening of emission during the pulse-on phase region as is also observed for IGR J17544$-$2619 in the lower panel of Fig. \ref{pfold}. 

The detection reported here, combined with past observations by other observatories, indicates that the pulsation signal produced by \object{IGR J17544$-$2619} is not always observable. \object{IGR J17544$-$2619} has had snapshot observations taken by \emph{XMM-Newton} (three $\sim$10\,ks exposures, \citealt{2004A&A...420..589G}) and \emph{Chandra} (19.6\,ks, \citealt{2005A&A...441L...1I}) along with a monitoring campaign by \emph{Swift}/XRT \citep{2008ApJ...687.1230S}, none of which show detections of the 71.49\,s signal. However the observational properties of SFXTs make the detection of pulse periods difficult. The biggest obstacle is the X-ray flaring time scale observed at soft X-ray energies, that is similar to likely pulsation periods (i.e. tens to hundreds of seconds). Flares dominate many of the soft X-ray data sets and mask pulsation signals due to the larger flux variations they induce. Many of the remaining data sets that are not dominated by flares are short \emph{Swift}/XRT exposures that have too short baselines or insufficient statistics for accurate timing analysis to be performed (e.g. Sidoli et al 2008). Given the nature of the previous soft X-ray observations of this source it is plausible that the pulsation of IGR J17544$-$2619 has not been detected prior to these observations, which show a steady flux at the 10$^{34}$ erg s$^{-1}$ level. These fluxes are consistent with those during which previous SFXT pulsations have been detected (e.g. \citealt{2009MNRAS.399..744G}).
 
Furthermore, the observation was performed as the neutron star was approaching apastron in the system which could also help explain why the pulsations were detected here. \citet{2009MNRAS.399L.113C} showed that at this orbital phase there is still a non-zero probability of stellar wind clump interaction in this system (\citealt{2009MNRAS.399L.113C}, Fig. 7). Additionally under the model of \citet{2009MNRAS.398.2152D} the stellar wind clumps expand as they move out from the companion supergiant star, likely becoming more homogeneous in density as they travel. As a result we may expect that the X-ray emission generated during the interaction of the neutron star and the expanded stellar wind clump would also be smoother and less prone to undergo the fast flares that dominate other soft X-ray observations of this source. Until a larger number of detections have been achieved however, formal conclusions on any link between the detection of the pulsations and the orbital phase of the observations cannot be drawn.        

\begin{figure}
	\includegraphics[scale=0.5]{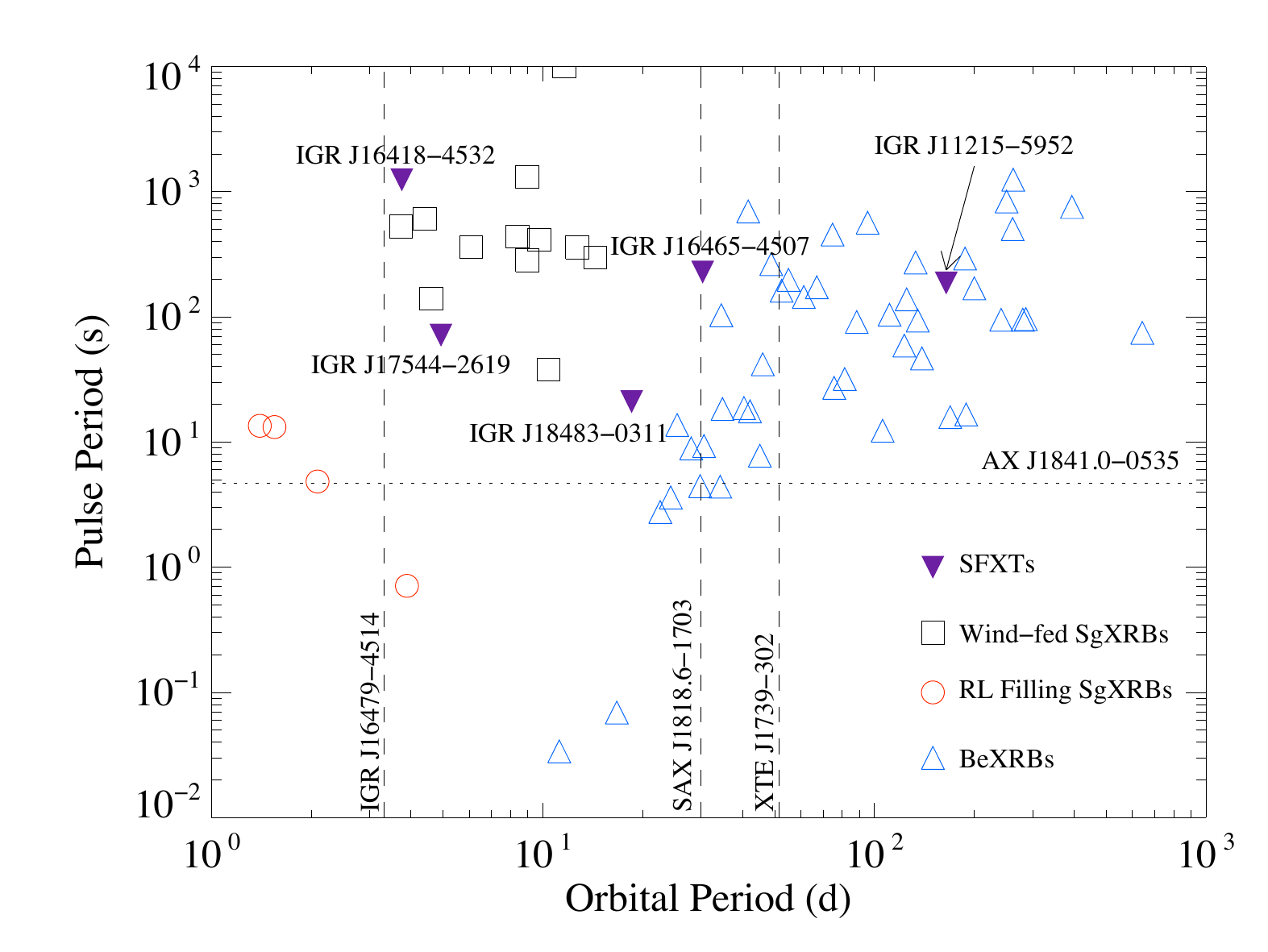}
	\caption{The Corbet Diagram showing the locations of the SFXT systems with at least one known period (Orbital or Pulse) \citep{1986MNRAS.220.1047C}. IGR J17544$-$2619 lies in the region of parameter space populated by the classical SgXRBs. }
	\label{corbet}
\end{figure}

Combining this detected pulsation with the 4.926\,d orbital period of \citet{2009MNRAS.399L.113C} allows the placement of IGR J17544$-$2619 on the Corbet diagram \citep{1986MNRAS.220.1047C}, Fig. \ref{corbet}. We see that IGR J17544$-$2619 is located close to the classical, wind-fed SgXRBs. Under the `clumpy wind' model of SFXTs \citep{2005A&A...441L...1I} the difference in behaviour seen when compared to the classical systems is explained by an enhanced eccentricity which results in the compact object spending only a fraction of its orbit within a dense stellar wind environment \citep{2007A&A...476..335W}. In this respect SFXTs can be considered as an extension of the classical SgXRBs that result from varying orbital parameters. However, some SFXTs have longer orbits and show orbital emission profiles that could be explained by the presence of a disk-like structure within the stellar wind of the supergiant companion, for example \object{IGR J11215$-$5952} \citep{2007A&A...476.1307S} and \object{XTE J1739$-$302} \citep{2010MNRAS.409.1220D}. Such characteristics are more akin to the BeXRB class of HMXBs. Currently the SFXT class is split into classic and intermediate SFXTs via the X-ray luminosity dynamic range observed from the system: $>$ 100 for an `intermediate' and $>$ 1000 for a `classic' SFXT. It is now becoming apparent that a distinction can also be drawn from the level of similarity of individual SFXTs to each of the classic HMXB family members and such similarities may be visualised by the Corbet diagram. This distinction, and the level to which some systems may again be `intermediate' between the two classes, should become more apparent as the number of SFXTs that can be placed on the Corbet diagram increases. Such a distinction could be an indication of a variety of stellar wind geometries present in SFXT systems and/or of varying evolutionary paths followed in their creation (see \citealt{2011MNRAS.415.3349L} for further details). 

\label{discuss}

\section{Conclusions}

Using observations from \emph{RXTE} we have detected a transient 71.49$\pm$0.02\,s signal that is most likely from the spinning neutron star in the SFXT IGR J17544$-$2619. The phase-folded light curve shows a double peaked structure where the emission is observed to harden during the peaks in the profile. The source was observed in a steady state of emission with a luminosity of $\sim$1$\times$10$^{34}$\,erg s$^{-1}$ (3$-$10\,keV). This pulse period places IGR J17544$-$2619 in the same region as the classical wind-fed SgXRBs on the Corbet diagram. The candidate SFXT IGR J16418$-$4532 also occupies this region, however other systems appear to be in intermediate (e.g. IGR J18483$-$0311) and BeXRB like positions (e.g. IGR J11215$-$5952) suggesting that SFXTs either span the gap between the classical types of HMXB or represent extreme examples of the known classes. A greater population of SFXTs is required on the Corbet diagram to investigate this further. We encourage further observations of the IGR J17544$-$2619 system using focusing X-ray telescopes to remove the current uncertainty in the origin of the detected pulsations and definitively prove that this is the pulsation of IGR J17544$-$2619. Phase-targeted, high-cadence observations of the source will also allow for a better understanding of the physical processes that could be causing the transient nature of the pulsation detection. Further observations of all SFXTs with one known periodicity (pulsation or orbital) are also vital to allow the largest possible sample to be placed on the Corbet diagram, furthering our knowledge of the nature of the class as whole.  

\label{conc}

\section*{Acknowledgements}

The authors wish to thank the anonymous referee for their helpful comments and suggestions.
The authors wish to thank R. H. D. Corbet and the PCA instrument team for their discussions on the time scales of systematic effects within the PCA background model. 
The authors also wish to thank J. J. M. in't Zand for his helpful discussion on the Galactic Ridge emission. 

S. P. Drave acknowledge support from the Science and Technology Facilities Council, STFC. L. J. Townsend is supported by a Mayflower scholarship from the University of Southampton.  A. Bazzano and V. Sguera acknowledge support from ASI/INAF contract n.I/009/10/0. This research has made use of the SIMBAD database, operated at CDS, Strasbourg, France. This research has made use of the IGR Sources page maintained by J. Rodriguez \& A. Bodaghee (http://irfu.cea.fr/Sap/IGR-Sources/).

\bibliographystyle{aa}
\bibliography{IGRJ17544_pulsations}

\end{document}